\documentclass{elsart}

\usepackage{epsfig}

\usepackage{amssymb}

\begin{document}

\begin{frontmatter}


  
  \begin{flushright}
    Cavendish-HEP-2003/24\\
    DAMTP-2003-94
  \end{flushright}

  \title{The Gluon Green's Function in the BFKL Approach at Next--to--Leading
    Logarithmic Accuracy}


\author[CAVENDISH,DAMTP]{Jeppe R.~Andersen}
\ead{andersen@hep.phy.cam.ac.uk},
\author[CAVENDISH]{Agust{\'\i}n Sabio Vera}
\ead{sabio@hep.phy.cam.ac.uk}

\address[CAVENDISH]{Cavendish Laboratory, University of Cambridge, Madingley Road, CB3 0HE, Cambridge, UK}
\address[DAMTP]{DAMTP, Centre for Mathematical Sciences, Wilberforce Road, CB3 0WA, Cambridge, UK}

\begin{abstract}
  We investigate the gluon Green's function in the high energy limit of QCD 
  using a recently
  proposed iterative solution of the Balitsky--Fadin--Kuraev--Lipatov (BFKL) 
  equation at next--to--leading logarithmic (NLL) accuracy. To establish the
  applicability of this method in the NLL approximation we solve the 
  BFKL equation as originally written by Fadin and Lipatov, and 
  compare the results with previous studies in the leading logarithmic (LL) 
  approximation.  
\end{abstract}


\end{frontmatter}

\section{Introduction}
\label{sec:introduction}
The Balitsky--Fadin--Kuraev--Lipatov \cite{FKL} formalism resums 
a class of logarithms dominant in the Regge limit of scattering amplitudes 
where the centre of mass energy $\sqrt{s}$ is large and the momentum transfer 
$\sqrt{-t}$ is fixed. Within this approach the high energy
cross--section for the process $A+B \rightarrow A'+B'$ can be written as 
\begin{eqnarray}
\label{cross--section1}
\sigma(s) &=&\int 
\frac{d^2 {\bf k}_a}{2 \pi{\bf k}_a^2}
\int \frac{d^2 {\bf k}_b}{2 \pi {\bf k}_b^2} ~\Phi_A({\bf k}_a) ~\Phi_B({\bf k}_b)
~f \left({\bf k}_a,{\bf k}_b, \Delta \equiv \ln{\frac{s}{s_0}}\right),
\end{eqnarray}
with $\Phi_{A,B}$ being the process--dependent impact factors and 
$f\left({\bf k}_a,{\bf k}_b,\Delta\right)$ the process--independent gluon 
Green's function. This gluon Green's function describes the interaction 
between two Reggeised gluons exchanged in
the $t$--channel with transverse momenta ${\bf k}_{a,b}$, and it carries 
the energy dependence of the cross--section. We choose to work with 
the symmetric Regge scale 
$s_0 = k_a k_b$, where $k_i \equiv \left|{\bf k}_i\right|$.

The resummation of terms of the form $\left(\alpha_s \Delta \right)^n$
defines the LL accuracy while the inclusion of contributions proportional to
$\alpha_s \left(\alpha_s\Delta \right)^n$ leads to the NLL approximation.
The ladder structure of the scattering amplitudes derived in the BFKL
formalism is described by an integral equation for the gluon Green's
function. The eigenfunctions of this integral equation are known at LL
accuracy and it is therefore possible to fully reconstruct the solution.  One
of the motivations to include the NLL contributions is to introduce running
coupling effects.  The logarithmic dependence introduced by the running
coupling terms in the NLL approximation significantly complicates the study
of the equation~\cite{running}.  During the past few years different
strategies have been suggested to study the NLL BFKL Green's
function.  In a fixed coupling analysis in  
Ref.~\cite{Douglas} it was first highlighted 
that the NLL corrections are large and negative compared to the LL. 
Different approaches to 
improve the convergence of the series expansion at NLL accuracy have been 
proposed~\cite{NLLpapers}. When running coupling effects are taken into 
account the situation improves as it has been shown in Ref.~\cite{CCSS}, in 
particular when those terms proportional to $\beta_0$ are resummed into 
$\alpha_s$. 

Recently, we proposed~\cite{Andersen:2003an} to use an iterative approach to
solve the equation in the NLL approximation.  A similar method was first
suggested at the LL accuracy in Ref.~\cite{LLAite}, where it was shown to
reproduce the analytic solution, and opened the possibility to perform
detailed LL phenomenological studies~\cite{LLenergy}.

The NLL formalism of Ref.~\cite{Andersen:2003an} has the advantage of dealing
with the BFKL kernel as calculated in
Ref.~\cite{Fadin:1998py,Ciafaloni:1998gs} with no approximations. In
particular, the solution includes all running coupling effects and, as we do
not use the angular averaged kernel, it allows for a complete study of
angular dependences.  The main purpose of this paper is to establish the
applicability of the NLL solution put forward in Ref.
\cite{Andersen:2003an}.  With such intention we will solve the NLL BFKL
equation as originally written in~\cite{Fadin:1998py}, with a particular
choice of renormalisation scale. Running coupling effects therefore
correspond to an expansion of the one--loop running of $\alpha_s(\mu)$ in the
${\overline{\rm MS}}$ scheme. A study of different schemes for the running of
the coupling and choices for the renormalisation scale will be presented in a
future publication.

This paper is organised as follows: In Section~\ref{sec:solution} we present
the BFKL equation and sketch the derivation of the iterative solution 
following Ref.~\cite{Andersen:2003an}. 
In Section~\ref{sec:numer-analys-nll} we present a numerical analysis 
of the NLL kernel, briefly indicating the mathematical expressions used in 
our numerical implementation.  
In Section~\ref{sec:Results} we use this numerical
implementation to study the NLL gluon Green's 
function. We analyse the convergence of the solution and
its dependence on the transverse momenta of the gluons exchanged in 
the $t$--channel. We present results on the evolution of the gluon Green's
function and a study of the renormalisation 
scale dependence.  We also show results on angular
dependences and a toy cross 
section obtained using simplified LL impact factors, to finally present 
our conclusions in Section 5.

\section{The Solution of the NLL BFKL Equation}
\label{sec:solution}
In this Section we sketch the method of solution for the BFKL equation at  
NLL accuracy proposed in Ref.~\cite{Andersen:2003an}. To write the equation in 
a convenient way we first perform a Mellin transform of the gluon Green's 
function, i.e.
\begin{equation}
\label{Mellin}
f \left({\bf k}_a,{\bf k}_b, \Delta\right) 
= \frac{1}{2 \pi i}
\int_{a-i \infty}^{a+i \infty} d\omega ~ e^{\omega \Delta} f_{\omega} 
\left({\bf k}_a ,{\bf k}_b\right).
\end{equation}
With such a transformation the BFKL equation in dimensional regularisation 
$\left(D = 4+2\epsilon \right)$ reads~\cite{Fadin:1998py}
\begin{eqnarray}
\label{eq:NLLBFKL}\omega f_\omega \left({\bf k}_a,{\bf k}_b\right) &=& \delta^{(2+2\epsilon)} 
\left({\bf k}_a-{\bf k}_b\right) + \int d^{2+2\epsilon}{\bf k}' ~
\mathcal{K}\left({\bf k}_a,{\bf k}'\right)f_\omega \left({\bf k}',{\bf k}_b 
\right),
\end{eqnarray}
where the kernel is expressed in terms of the gluon Regge trajectory  and 
the real emission component in the following way
\begin{eqnarray}
\mathcal{K}\left({\bf k}_a,{\bf k}\right) = 2 \,\omega^{(\epsilon)}\left({\bf k}_a^2\right) \,\delta^{(2+2\epsilon)}\left({\bf k}_a-{\bf k}\right) + \mathcal{K}^{(\epsilon)}_r\left({\bf k}_a,{\bf k}\right) + {\tilde \mathcal{K}}_r\left({\bf k}_a,{\bf k}\right).
\end{eqnarray}
Both $\epsilon$--dependent parts of the kernel contain $1/\epsilon$ and $1/\epsilon^2$ poles. The cancellation of the poles in the trajectory against those in the real emission kernel was shown in Ref.~\cite{Andersen:2003an} by introducing a phase space slicing parameter $\lambda$. The BFKL equation can then be expressed as
\begin{eqnarray}
\label{nll}
\left(\omega - \omega_0\left({\bf k}_a^2,\lambda^2\right)\right) f_\omega \left({\bf k}_a,{\bf k}_b\right) &=& \delta^{(2)} \left({\bf k}_a-{\bf k}_b\right)\\
&&\hspace{-5cm}+ \int d^2 {\bf k} \left(\frac{1}{\pi {\bf k}^2} \xi \left({\bf k}^2\right) \theta\left({\bf k}^2-\lambda^2\right)+\widetilde{\mathcal{K}}_r \left({\bf k}_a,{\bf k}_a+{\bf k}\right)\right)f_\omega \left({\bf k}_a+{\bf k},{\bf k}_b\right),\nonumber
\end{eqnarray}
where $\widetilde{\mathcal{K}}_r$ is the finite part of the
emission kernel, 
\begin{eqnarray}
\label{non_ang_av}
\widetilde{\mathcal{K}}_r \left({\bf q},{\bf q}'\right) &=& 
\frac{\bar{\alpha}_s^2(\mu)}{4 \pi} 
\left\{\left(1+\frac{n_f}{N_c^3}\right)
\frac{\left(3({\bf q}\cdot{\bf q'})^2
-2 {\bf q}^2 {\bf q'}^2 \right)}{16 {\bf q}^2 {\bf q'}^2}
\left(\frac{2}{{\bf q}^2}+\frac{2}{{\bf q'}^2} \right.\right. \nonumber\\
&&\hspace{-2cm}\left.+\left(\frac{1}{{\bf q'}^2}-\frac{1}{{\bf q}^2}\right)
\ln{\frac{{\bf q}^2}{{\bf q'}^2}}\right) 
+\frac{2({\bf q}^2-{\bf q'}^2)}{({\bf q}-{\bf q'})^2({\bf q}+{\bf q'})^2} 
\left(\frac{1}{2}\ln{\frac{{\bf q}^2}{{\bf q'}^2}}
\ln{\frac{{\bf q}^2 {\bf q'}^2 ({\bf q}-{\bf q'})^4}
{({\bf q}^2+{\bf q'}^2)^4}} \right.  \nonumber\\
&&\left.+ \left( \int_0^{- {\bf q}^2 / {\bf q'}^2} -
\int_0^{- {\bf q'}^2 / {\bf q}^2} \right) 
dt \frac{\ln(1-t)}{t}\right)-\frac{1}{({\bf q}-{\bf q'})^2}\ln^2{\frac{{\bf q}^2}{{\bf q'}^2}}\nonumber\\
&&\hspace{-2cm}-\left(3+\left(1+\frac{n_f}{N^3_c}\right)
\left(1-\frac{({\bf q}^2+{\bf q'}^2)^2}{8 {\bf q}^2 {\bf q'}^2}
- \frac{(2 {\bf q}^2 {\bf q'}^2 - 3 {\bf q}^4 - 3 {\bf q'}^4)}
{16 {\bf q}^4 {\bf q'}^4}({\bf q} \cdot {\bf q'})^2\right)\right)\nonumber\\
&&\hspace{1cm}\times 
\int^\infty_0 dx \frac{1}{{\bf q}^2 + x^2 {\bf q'}^2} \ln{\left|\frac{1+x}{1-x}\right|}\nonumber\\
&&\hspace{-2cm}\left.-\left(1-\frac{({\bf q}^2-{\bf q'}^2)^2}{({\bf q}-{\bf q'})^2
({\bf q}+{\bf q'})^2}\right) 
\left( \left( \int_0^1 
-\int_1^\infty \right) dz \frac{1}{({\bf q'}-z {\bf q})^2}
\ln{\frac{(z {\bf q})^2}{{\bf q'}^2}}\right)\right\}.
\end{eqnarray}

The function 
\begin{eqnarray}
\xi \left({\rm X}\right) &=& \bar{\alpha}_s(\mu) +  
\frac{{\bar{\alpha}_s}^2(\mu)}{4}\left[\frac{4}{3}-\frac{\pi^2}{3}+\frac{5}{3}\frac{\beta_0}{N_c}-\frac{\beta_0}{N_c}\ln{\frac{{\rm X}}{\mu^2}}\right],
\label{xi}
\end{eqnarray}
with $\bar{\alpha}_s(\mu) \equiv \alpha_s (\mu) N_c / \pi$, plays a crucial role
in the cancellation of infrared singularities in the final result. It
contains the information about the running coupling effects encoded in the
terms proportional to $\beta_0 \equiv \frac{11}{3}N_c-\frac{2}{3}n_f$. This
$\xi$ function can be modified to resum these effects and consider them in
different schemes. We will study this in a separate publication. In the
present work we take $\xi$ as in Eq.~(\ref{xi}), which corresponds to the
one--loop expansion of $\alpha_s(\mu)$ in the ${\overline{\rm MS}}$ scheme with
renormalisation scale $\mu$. In this approach the expression for the gluon
Regge trajectory reads
\begin{eqnarray}
\label{eq:omega0}
\omega_0 \left({\bf q}^2,\lambda^2 \right) &=& - \xi\left(\left|{\bf q}\right|\lambda\right) \ln{\frac{{\bf q}^2}{\lambda^2}} + {\bar{\alpha}_s}^2(\mu) \frac{3}{2} \zeta (3).
\end{eqnarray}

As explained in Ref.~\cite{Andersen:2003an} one can solve the BFKL equation by 
iterating Eq.~(\ref{nll}) in such a way that an extra $\omega$--pole is  
generated per rung in the BFKL ladder. It is then 
possible to perform the inverse Mellin transform to find the solution 
directly in energy space, i.e.
\begin{eqnarray}
\label{ours}
f({\bf k}_a ,{\bf k}_b, \Delta) 
&=& \exp{\left(\omega_0 \left({\bf k}_a^2,{\lambda^2}\right) \Delta \right)}
\left\{\frac{}{}\delta^{(2)} ({\bf k}_a - {\bf k}_b) \right. \\
&&\hspace{-2cm}+ \sum_{n=1}^{\infty} \prod_{i=1}^{n} 
\int d^2 {\bf k}_i \left[\frac{\theta\left({\bf k}_i^2 - \lambda^2\right)}{\pi {\bf k}_i^2} \xi\left({\bf k}_i^2\right) + \widetilde{\mathcal{K}}_r \left({\bf k}_a+\sum_{l=0}^{i-1}{\bf k}_l,
{\bf k}_a+\sum_{l=1}^{i}{\bf k}_l\right)\frac{}{}\right]\nonumber\\
&& \hspace{-1cm} \times  
\int_0^{y_{i-1}} d y_i ~ {\rm exp}\left[\left(
\omega_0\left(\left({\bf k}_a+\sum_{l=1}^i {\bf k}_l\right)^2,\lambda^2
\right)\right.\right.\nonumber\\
&&\left.\left.\left.\hspace{0cm}
-\omega_0\left(\left({\bf k}_a+\sum_{l=1}^{i-1} {\bf k}_l\right)^2,
{\lambda^2}\right)\right) y_i\right] \delta^{(2)} \left(\sum_{l=1}^{n}{\bf k}_l 
+ {\bf k}_a - {\bf k}_b \right)\right\}, \nonumber
\end{eqnarray}
where we have used the notation $y_0\equiv\Delta$. 

In the numerical implementation discussed in this work we have chosen the
renormalisation scale to be $\mu = k_b$, one of the perturbative scales in
the interaction.  The study of alternatives to this choice will be presented
elsewhere.

It is important to realise that Eq.~(\ref{ours}) gives the correct solution
in the $\lambda \rightarrow 0$ limit. In practice we are able to numerically
check the region of stability of the expression in Eq.~(\ref{ours}), i.e. the
region at small $\lambda$ where the result is flat in $\lambda$. Every
extra term in the series expansion corresponds to an additional iteration of
the kernel in the integral equation. For a given value of the variable
$\Delta$ and the slicing parameter $\lambda$ only a finite number of terms in
the expansion is needed to obtain the solution to a given accuracy.

In the next Section we show the mathematical expressions used for the kernel
in our implementation, and briefly describe the structure of the trajectory
and the real emission kernel.

\section{Analysis of the BFKL Trajectory and Kernel}
\label{sec:numer-analys-nll}
\begin{figure} 
\centerline{\psfig{file=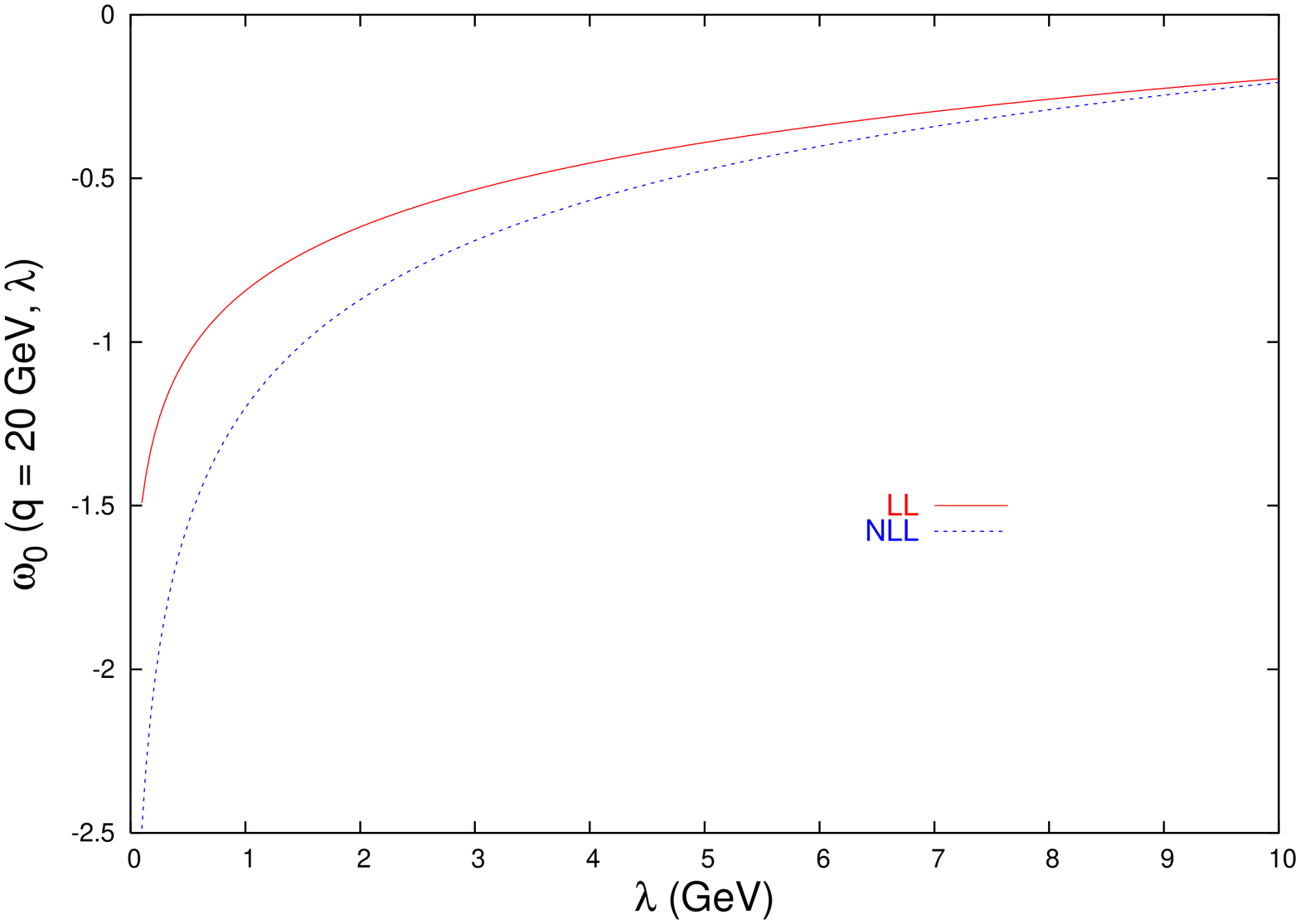,width=4.8cm,angle=-90}\psfig{file=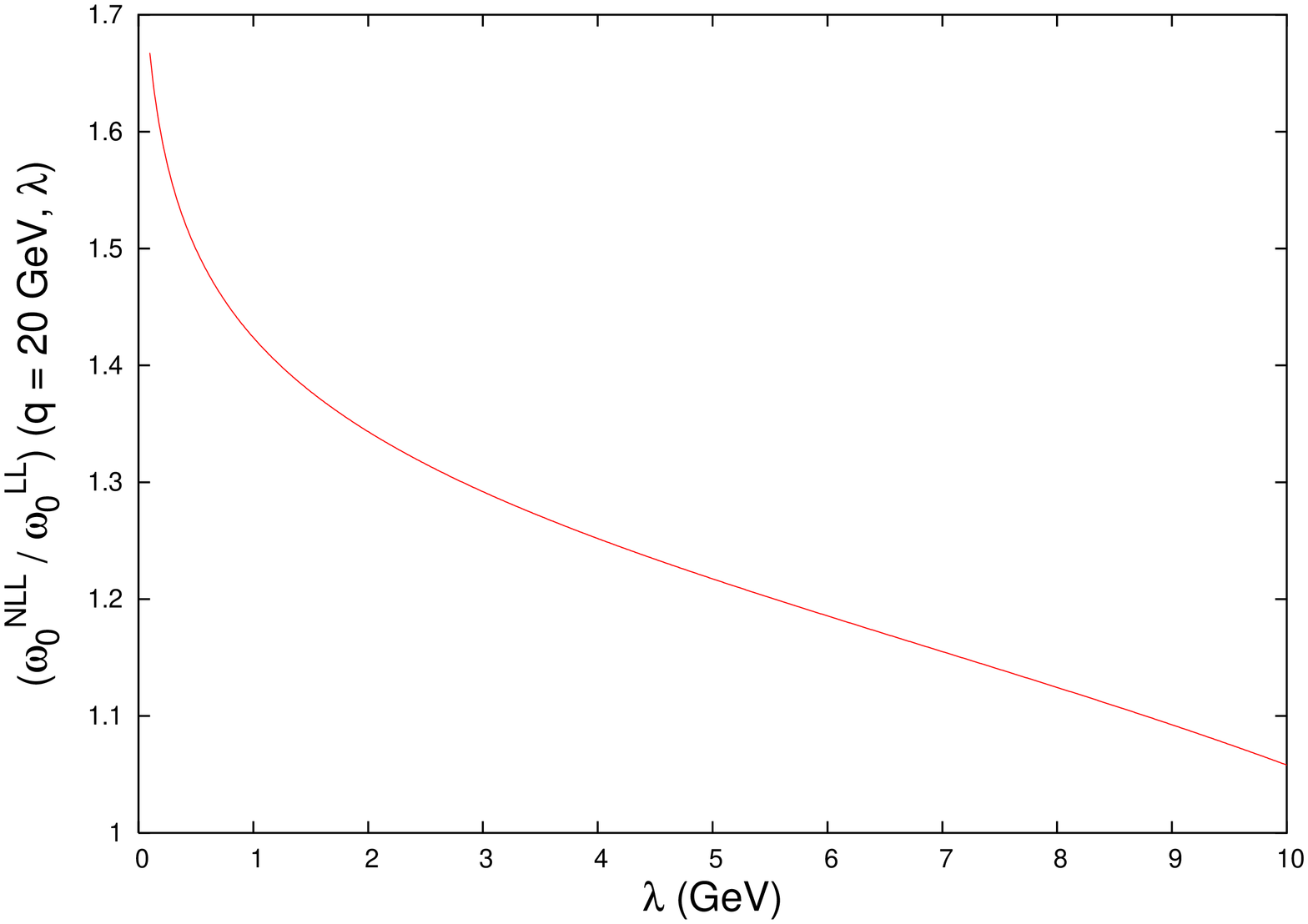,width=4.8cm,angle=-90}}
\centerline{\psfig{file=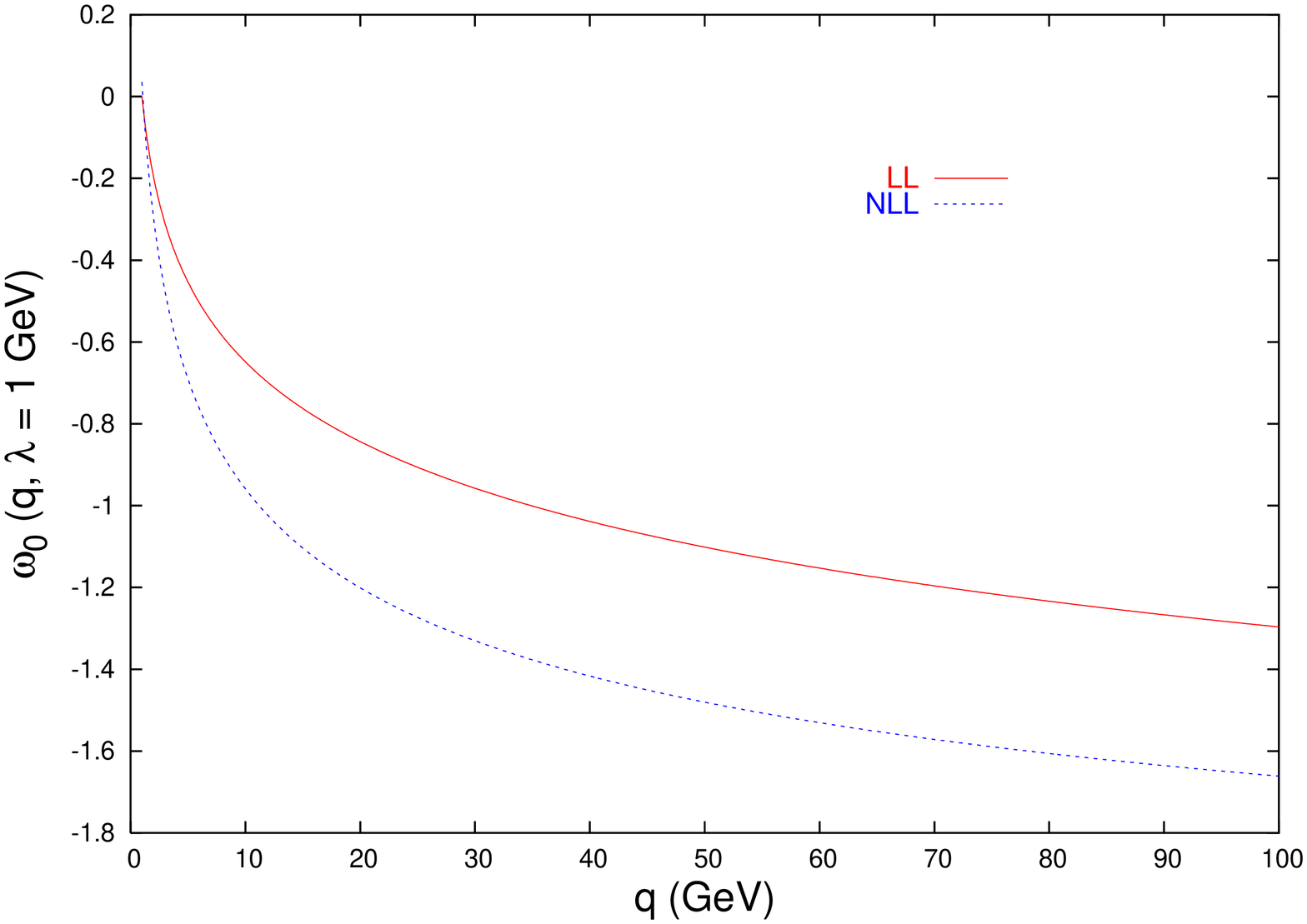,width=4.8cm,angle=-90}\psfig{file=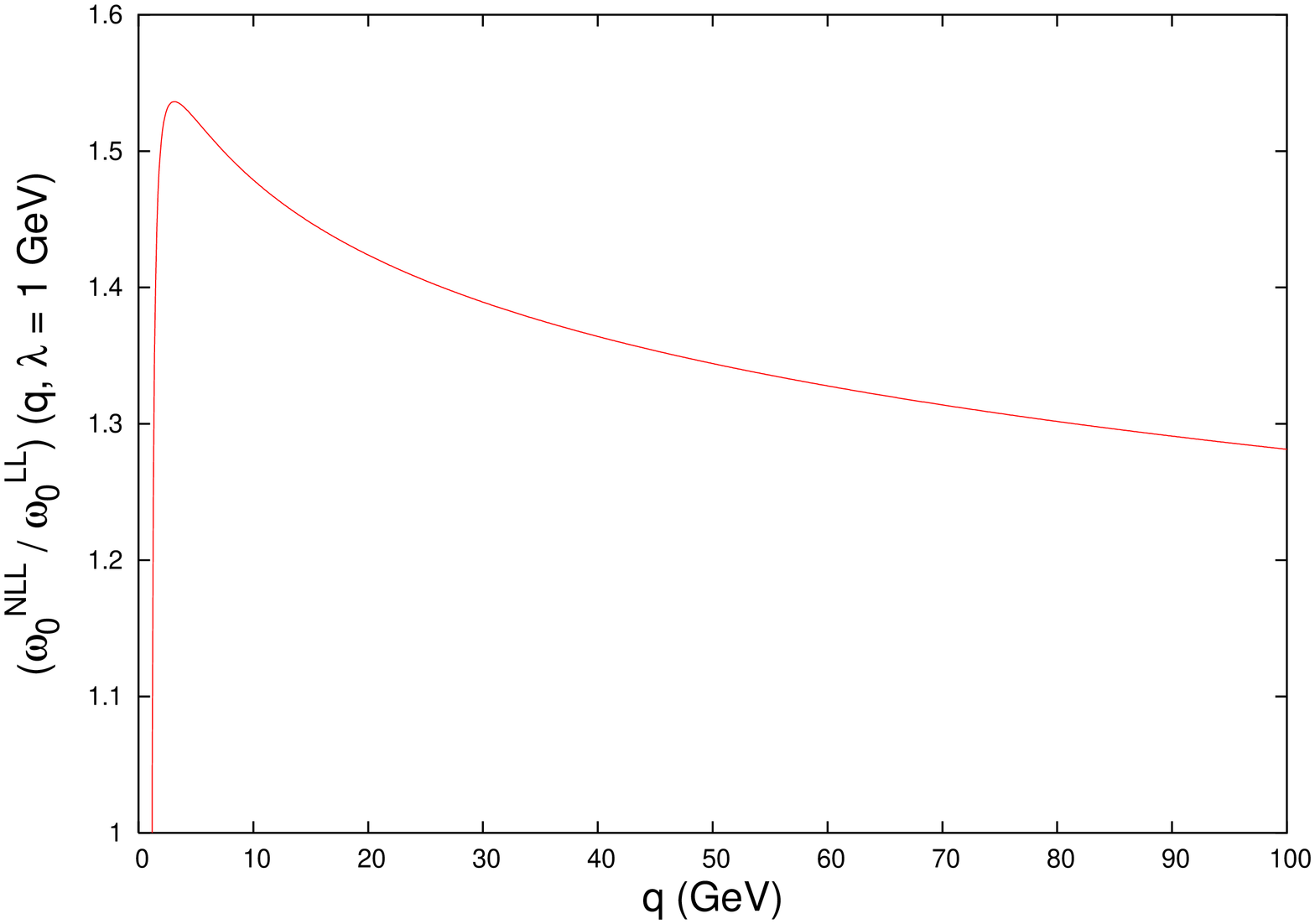,width=4.8cm,angle=-90}}
\centerline{\psfig{file=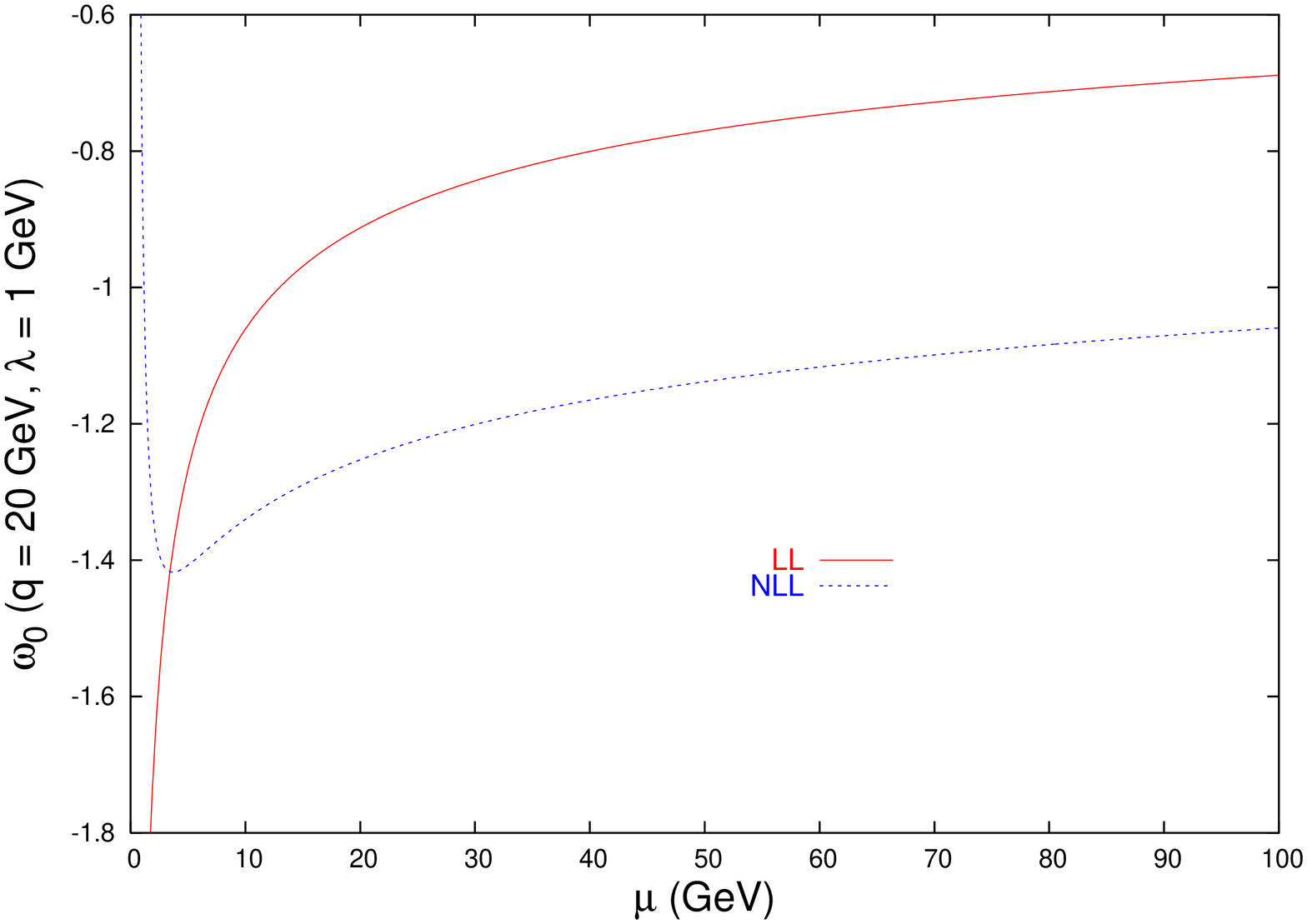,width=4.8cm,angle=-90}\psfig{file=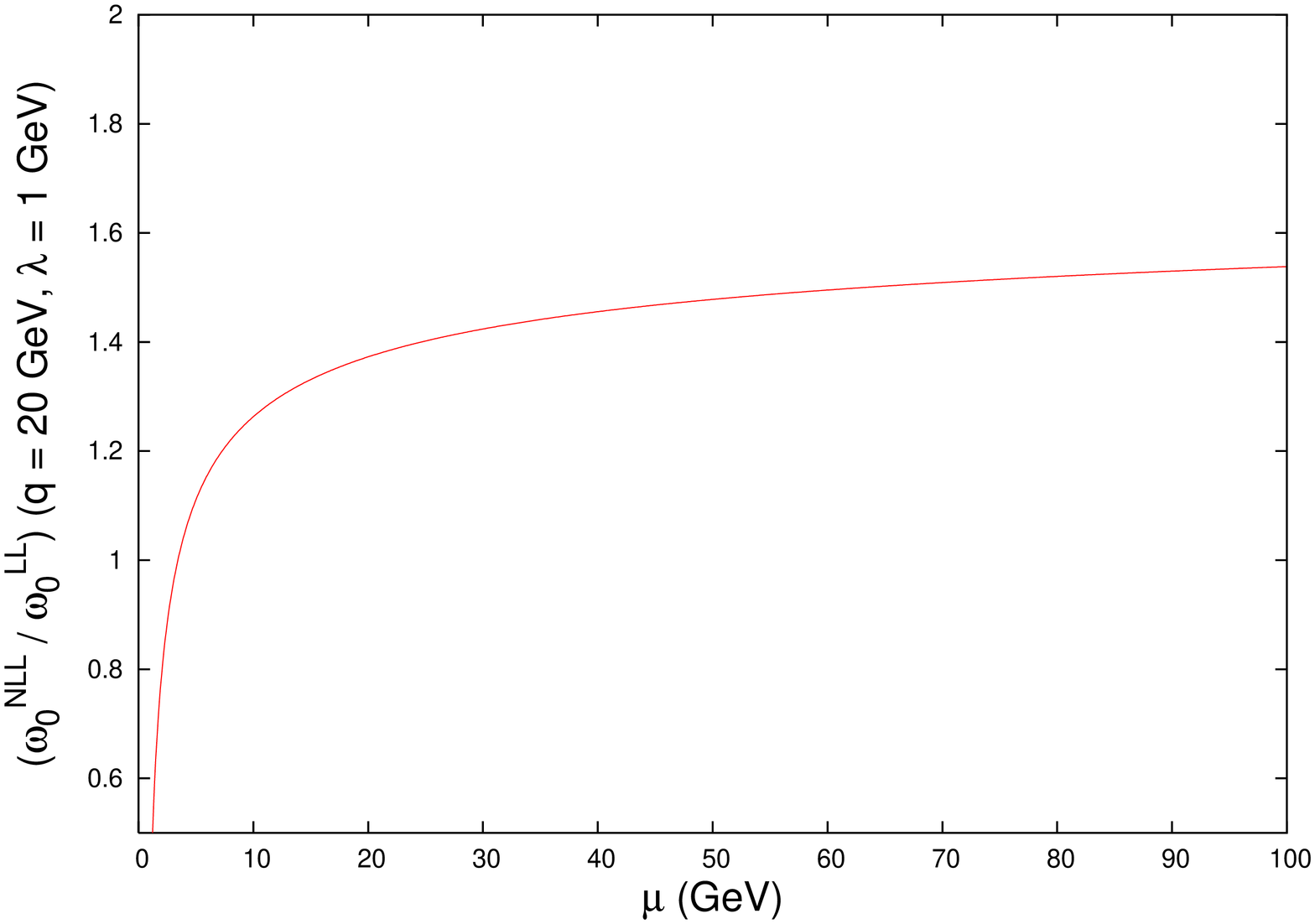,width=4.8cm,angle=-90}}
\caption{Comparison of the gluon Regge trajectory calculated at LL   
and at NLL.}
\label{trajectory}
\end{figure}

As we have already pointed out, there are two key elements in the BFKL
equation: the gluon Regge trajectory, $\omega_0 \left({\bf
    q}^2,\lambda^2\right)$, which contains the function $\xi$, and the finite
real emission component of the kernel, $\widetilde{\mathcal{K}}_r \left({\bf
    q},{\bf q}'\right)$.  The trajectory is not infrared--finite and
therefore carries a $\lambda$ dependence. It is interesting to compare the
behaviour of the trajectory at LL to that at NLL. This is shown in
Fig.~\ref{trajectory}, where first we display the dependence on $\lambda$
with $q = 20$ GeV and then, for a fixed value of $\lambda = 1$ GeV, we plot
the behaviour in $q$. The values used in these plots are $n_f = 4$, $N_c =
3$, and, for the top four graphs, $\mu = 30$~GeV. For a value of
$\Lambda^{(4)}_\mathrm{QCD}$ of 0.1416~GeV this implies that $\alpha_s(\mu) =
0.1408$.
For this value of the renormalisation scale the NLL trajectory is always more
negative than the LL one with the trajectories separating from each other as
the slicing parameter decreases for a fixed $q$, or, when we fix $\lambda$,
for large values of $q$. For fixed values of $q$ and $\lambda$, and for large
renormalisation scales the effect of changing $\mu$ does not affect the
difference between the trajectory at LL and NLL, as can be seen in the bottom
two plots in Fig 1. This is not the case for lower values of $\mu$, where the
trajectories can even overlap at the minimum of $\omega_0^{NLL}$.

\begin{figure} 
\centerline{\psfig{file=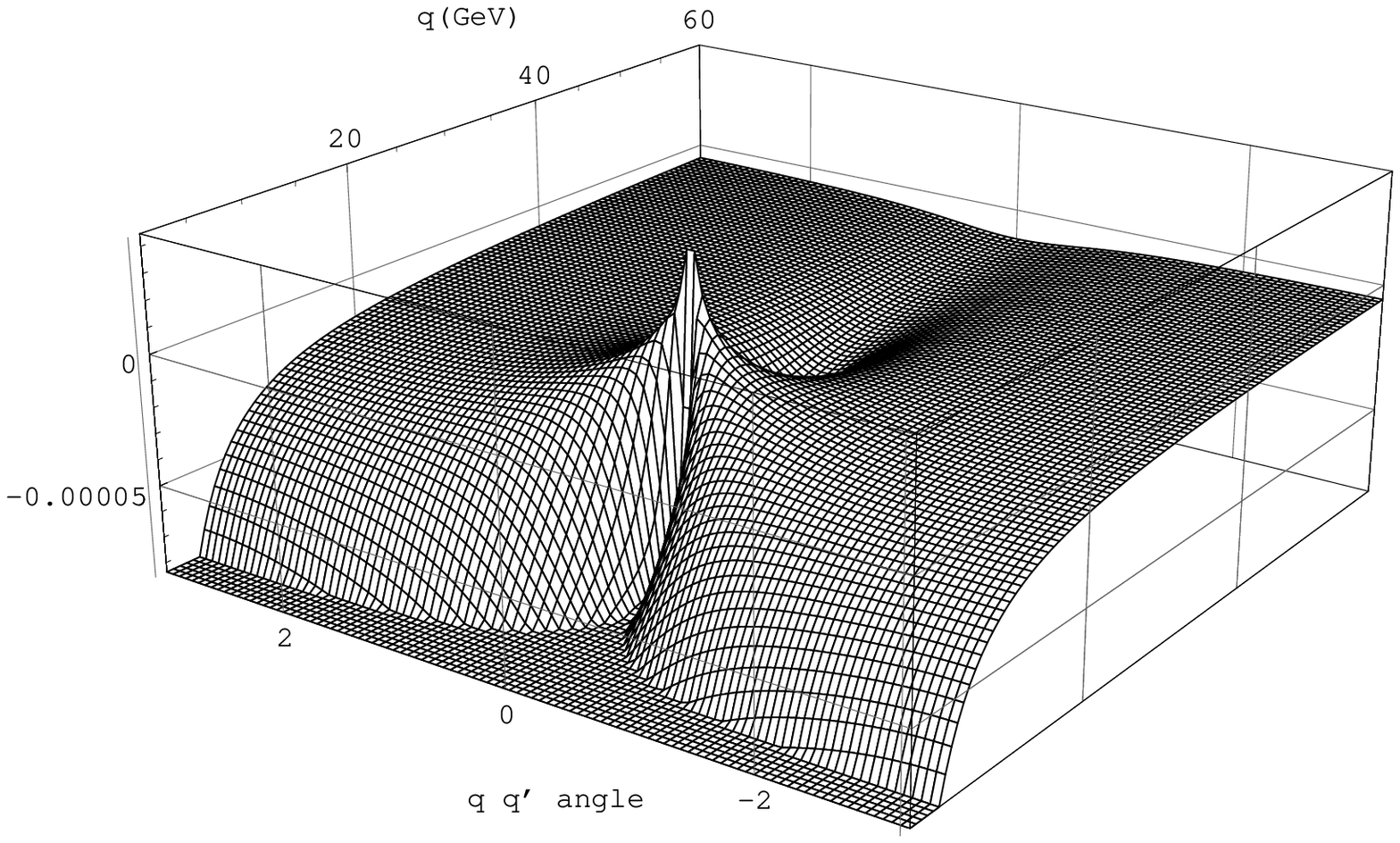,width=14cm,angle=0}}
\caption{Structure of the kernel $\tilde{K}_r \left(q,q',\theta\right)$ (in GeV$^{-2}$) for 
$q' = 20$ GeV as a function of $q$ and the angle between ${\bf q}$ and 
${\bf q'}$.}
\label{Kernel}
\end{figure}

To implement the solution for the gluon Green's function numerically, 
the kernel in Eq. (\ref{non_ang_av}) is rewritten as
\begin{eqnarray}
\label{non_ang_av_rewritten}
&&\hspace{-1.2cm}\widetilde{\mathcal{K}}_r \left(q,q',\theta\right) ~=~ 
\frac{\bar{\alpha}_s^2}{4 \pi} 
\left\{ \left(1+\frac{n_f}{N_c^3}\right)
\left(\frac{3 \cos^2{\theta}-2}{16}\right) 
\left(\frac{2}{q^2}+\frac{2}{q'^2}
+\left(\frac{1}{q'^2}-\frac{1}{q^2}\right)
\ln{\frac{q^2}{q'^2}}\right) \right.\nonumber\\
&&\hspace{-0.2cm}
+\frac{2(q^2-{q'}^2)}{\left( (q^2 + {q'}^2)^2-4\,q^2\,{q'}^2
\,\cos^2{\theta}\right)} 
\left(\frac{1}{2}\ln{\frac{q^2}{{q'}^2}}
\ln{\frac{{q}^2 {q'}^2 \left({q}^2+{q'}^2-2\,q\,q'\,\cos{\theta}\right)^2}
{({q}^2+{q'}^2)^4}} \right.\nonumber\\
&&\hspace{-0.2cm}\left. + \left( \int_0^{- {q}^2 / {q'}^2} -
\int_0^{- {q'}^2 / {q}^2} \right) 
dt \frac{\ln(1-t)}{t}\right) - 
\frac{1}{\left(q^2+{q'}^2-2\,q\,q'\,\cos{\theta}\right)}
\ln^2{\frac{q^2}{{q'}^2}} \nonumber\\
&&\hspace{-0.2cm}-\left(3+\left(1+\frac{n_f}{N^3_c}\right)
\left(1-\frac{({q}^2+{q'}^2)^2}{8 {q}^2 {q'}^2}
- \frac{(2 {q}^2 {q'}^2 - 3 {q}^4 - 3 {q'}^4)}
{16 {q}^2 {q'}^2} \cos^2{\theta}\right)\right)\nonumber\\
&&\hspace{4cm}\times 
\frac{1}{q\,q'}\left(\ln{\frac{q^2}{{q'}^2}}\arctan{\frac{q'}{q}}+
2\, {\rm Im}\left\{{\rm Li}_2 \left(i\,\frac{q'}{q}\right)\right\}\right)
\nonumber\\
&&\hspace{0cm}\left. -\frac{2\,q\,{q'}\,\left|\sin{\theta}\right|}{
\left(q^2-{q'}^2\right)^2+4\,q^2\,{q'}^2\,\sin^2{\theta}} 
\left(\mathcal{F}\left(q,q',\theta\right)
+\mathcal{F} \left(q',q,\theta\right)\right)\right\},
\end{eqnarray}
where $\theta$ is the angle between the two--dimensional vectors ${\bf q}$ and ${\bf q}'$, and
\begin{eqnarray}
&&\hspace{-1cm}\mathcal{F} \left(q,q',\theta\right) ~=~ 
{\rm Im} \left\{4\,{\rm Li}_2 \left(\frac{q}{q'}\,e^{-i \left|\theta\right|}\right)- \ln{\frac{q^2}{{q'}^2}}\ln{\frac{q'\,\left|\sin{\theta}\right|-i\left(q-q'\,\cos{\theta}\right)}{q'\,\left|\sin{\theta}\right|+i\left(q-q'\,\cos{\theta}\right)}}\right\}
\end{eqnarray}
with 
\begin{eqnarray}
{\rm Li}_2 \left(z\right) &=& - \int^z_0 dt \frac{\ln(1-t)}{t}.
\end{eqnarray}
This kernel has an integrable singularity in the ${\bf q} \rightarrow {\bf
  q}'$ limit.
As an example of its structure in the vicinity of this singularity we plot
$\widetilde{\mathcal{K}}_r \left(q,q'=20 ~{\rm GeV},\theta\right)$ in
Fig.~\ref{Kernel} with the same values for the parameters as those in
Fig.~\ref{trajectory} where the renormalisation scale was fixed to $\mu =
30$~GeV.

In the next Section we use this form of the kernel in order to study the 
solution to the NLL BFKL equation.

\section{Study of the Gluon's Green Function}
\label{sec:Results}
We have implemented Eq.~(\ref{ours}) in a Monte Carlo integration routine
using the form of the kernel presented in Section~\ref{sec:numer-analys-nll}.
In the following analysis, we have chosen to run the coupling at one loop,
matching the values of Ref.~\cite{Martin:1999ww} and, as previously
explained, we use $\mu = k_b$ as the renormalisation scale and set $n_f=4$.
The function $\xi$ will always correspond to Eq.~(\ref{xi}).

\subsection{Convergence of the Solution}
\label{sec:convergence-solution}
Before presenting the results for the gluon Green's function, we first
investigate the properties of convergence for Eq.~(\ref{ours}). There are two
points of interest: Firstly, how the $\lambda\to 0$ limit is approached, and
secondly, to determine how many terms in the infinite sum of Eq.~(\ref{ours})
are needed to obtain the solution within a given numerical accuracy. These
two points are linked because the smaller the value of $\lambda$, the more
terms are necessary in the expansion to reach good accuracy, but also the
better the approximation $f_\omega \left({\bf k}_a+{\bf k},{\bf k}_b\right)
\simeq f_\omega \left({\bf k}_a,{\bf k}_b\right)$ for $\left|{\bf k}\right| <
\lambda$ (as shown in Ref.~\cite{Andersen:2003an} this approximation is used
in order to write the BFKL equation as in Eq.~(\ref{nll})). A good choice for
$\lambda$ is therefore characterised by being small enough for the
approximation to be valid, ensuring in this way an accurate answer, while
being large enough to warrant the contribution from only a finite number of
terms in the expansion, so that the numerical evaluation is fast.

\begin{figure}[tbp]
  \centering
  \epsfig{width=10cm,file=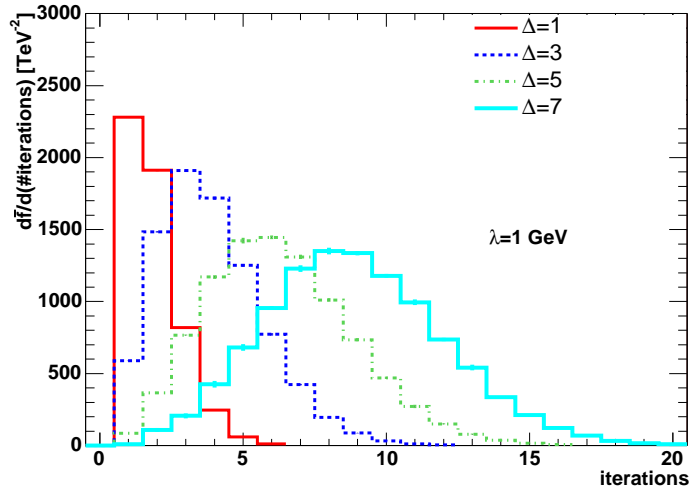}
  \caption{Distribution on the number of iterations in building up the 
    NLL gluon Green's function for different values of the parameter $\Delta$,
    at a fixed value of $\lambda=1$~GeV. The gluon Green's function is
    evaluated for $k_a=25$~GeV, $k_b=30$~GeV and the renormalisation scale is
    chosen to be $\mu=k_b$.}
  \label{fig:ngplot}
\end{figure}

In Fig.~\ref{fig:ngplot} we plot the contribution from successive terms in
the infinite sum of Eq.~(\ref{ours}) to the angular integrated NLL gluon
Green's function
\begin{eqnarray}
\label{aver}
{\bar f} \left(k_a, k_b, \Delta \right) &=& \int_0^{2 \pi} 
d\theta \, f \left(k_a, k_b, \theta, \Delta \right) ,
\end{eqnarray}
with $\theta$ being the angle between $\mathbf{k}_a$ and $\mathbf{k}_b$, at
${k}_a = 25$~GeV, ${k}_b =30$~GeV, $\lambda=1$~GeV, and for different values
of the parameter $\Delta$.  These values have been chosen with no other
intention but to illustrate the capability of the NLL formalism proposed in
Ref.~\cite{Andersen:2003an} and its numerical implementation. We see that for
a given choice of parameters, only a finite number of terms contribute to the
infinite sum. All the results presented in this paper have been calculated
with some upper limit on the number of terms included in the infinite sum of
Eq.~(\ref{ours}). It has been verified that this limit is put sufficiently
high as to reproduce the solution with the required accuracy. This plot
contains information about how the emission builds up, revealing in a
quantitative way the fact that when the energy available for the scattering
process is larger, the distribution peaks at larger values of the number of
iterations. Although this trend is independent of $\lambda$, the specific
position of the peak is not.

$\bar f({k}_a,{k}_b,\Delta)$ in Eq.~(\ref{aver}) could also
have been obtained by first angular averaging the BFKL kernel and then
solving the equation. Although in our solution we have all the angular
information, in this case we average in angles in order to compare with the
analytic expression for the LL BFKL gluon Green's function with zero
conformal spin, i.e.
\begin{eqnarray}
\bar f \left({k}_a,{k}_b,\Delta\right) &=& \frac{4}{k_a k_b} 
\int_0^\infty d\nu \left(\frac{k_a^2}{k_b^2}\right)^{i \nu} 
e^{{\bar \alpha}_s \Delta {\chi_0} (\nu)}
\label{LLAna}
\end{eqnarray}
with the LL eigenvalue being
\begin{eqnarray}
\chi_0 (\nu) &=& -2 \,{\rm Re}\left\{\psi \left(\frac{1}{2}+i \nu
\right)-\psi (1)\right\}.
\end{eqnarray}
The LL limit of Eq.~(\ref{ours}) is given by
\begin{eqnarray}
f({\bf k}_a ,{\bf k}_b, \Delta) 
&=& \left(
\frac{\lambda^2}{k_a^2}\right)^{\bar{\alpha}_s \,\Delta}
\left\{  \frac{}{}\delta^{(2)} ({\bf k}_a - {\bf k}_b) 
+  \sum_{n=1}^{\infty} \prod_{i=1}^{n} 
\bar{\alpha}_s \int d^2{\bf k}_i 
\frac{\theta\left({\bf k}_i^2- \lambda^2\right)}{\pi {\bf k}_i^2} \right.\\
&\times& \left. \int_0^{y_{i-1}} d y_i 
\left(\frac{\left({\bf k}_a+\sum_{l=1}^{i-1} {\bf k}_l\right)^2}
{\left({\bf k}_a+\sum_{l=1}^i {\bf k}_l\right)^2}\right)
^{\bar{\alpha}_s \, y_i} \delta^{(2)} \left(\sum_{l=1}^{n}{\bf k}_l 
+ {\bf k}_a - {\bf k}_b \right) \right\}.\nonumber
\label{LLItera}
\end{eqnarray}
In this study we have checked that the LL analytic results from
Eq.~(\ref{LLAna}) coincide with those from the LL version of our numerical
implementation. This will be illustrated in the plots below.

\begin{figure}[tbp]
  \centering
  \epsfig{width=10cm,file=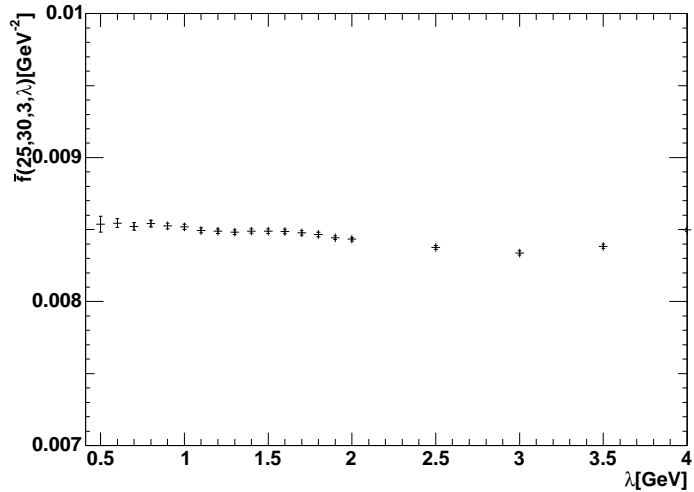}
  \caption{Dependence of the NLL solution for the gluon Green's function
    on the parameter $\lambda$ for $k_a=25$~GeV, $k_b=30$~GeV and
    $\Delta=3$.}
  \label{fig:lambda}
\end{figure}

Having shown how it can be checked that a sufficient number of iterations in
Eq.~(\ref{ours}) has been performed for a given choice of the parameters, we
will now proceed to study the $\lambda\to0$ limit of the equation. In
Fig.~\ref{fig:lambda} we have plotted the $\lambda$--dependence of the
angular integrated gluon Green's function on $\lambda$ for a choice of $k_a =
25$ GeV, $k_b = 30$ GeV and $\Delta=3$. The result is very flat in $\lambda$
for small values of this parameter, demonstrating, remarkably, the
cancellation between the infrared divergences present in the gluon NLL Regge
trajectory and those stemming from the integration of the real NLL emission
kernel over phase space. For larger values of $\lambda$ we observe a growing
$\lambda$--dependence originating from our initial approximation $f_\omega
\left({\bf k}_a+{\bf k},{\bf k}_b\right) \simeq f_\omega \left({\bf k}_a,{\bf
    k}_b\right)$ for $\left|{\bf k}\right| < \lambda$.  In all the results
presented in this work we have taken $\lambda=1$~GeV and checked that this
choice is in a region with a very weak dependence on $\lambda$.

\subsection{Dependence of the Gluon Green's Function on the External Momenta}
\label{sec:depend-gluon-greens}
\begin{figure}[tbp]
  \centering
  \epsfig{width=10cm,file=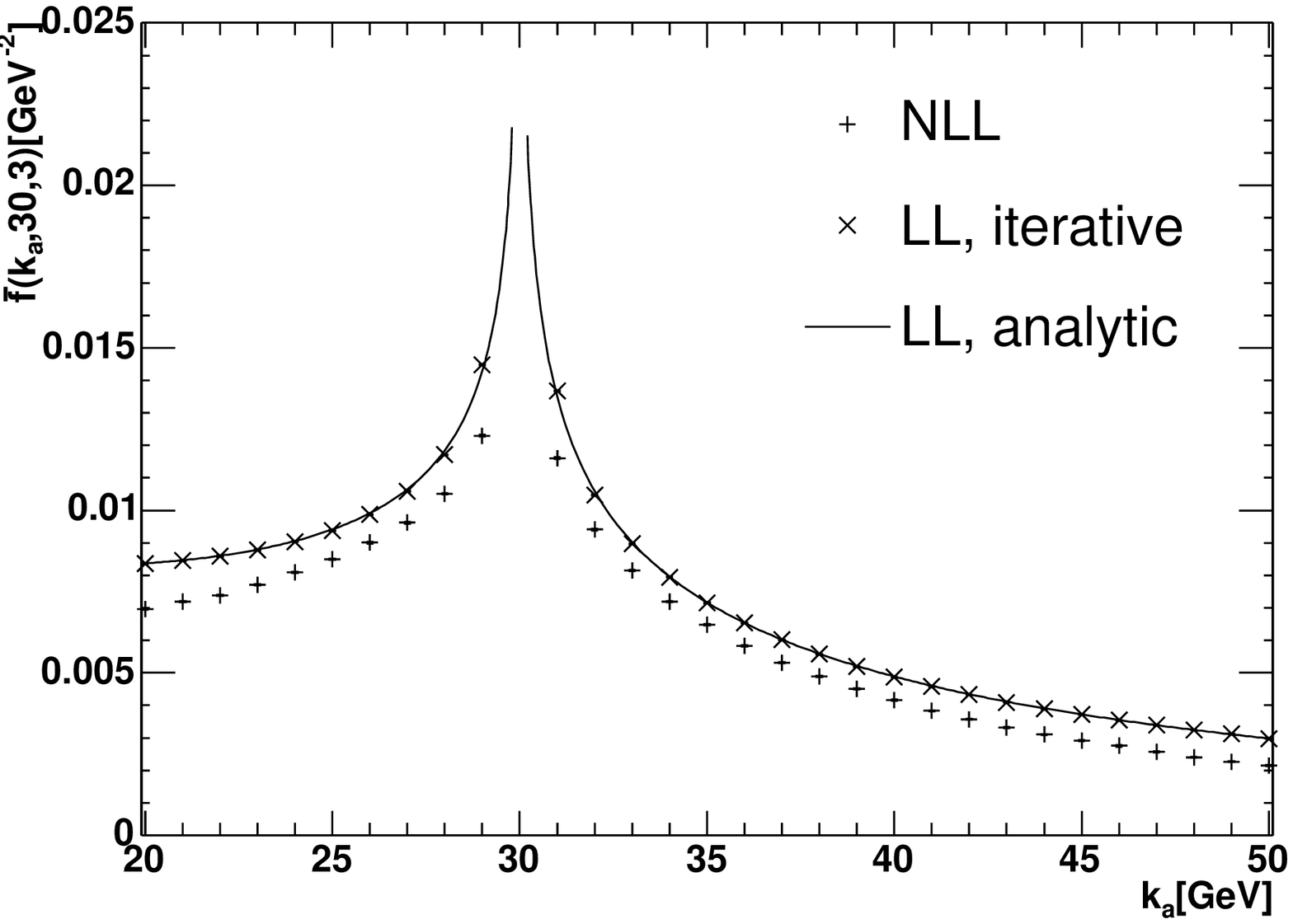}
  \epsfig{width=10cm,file=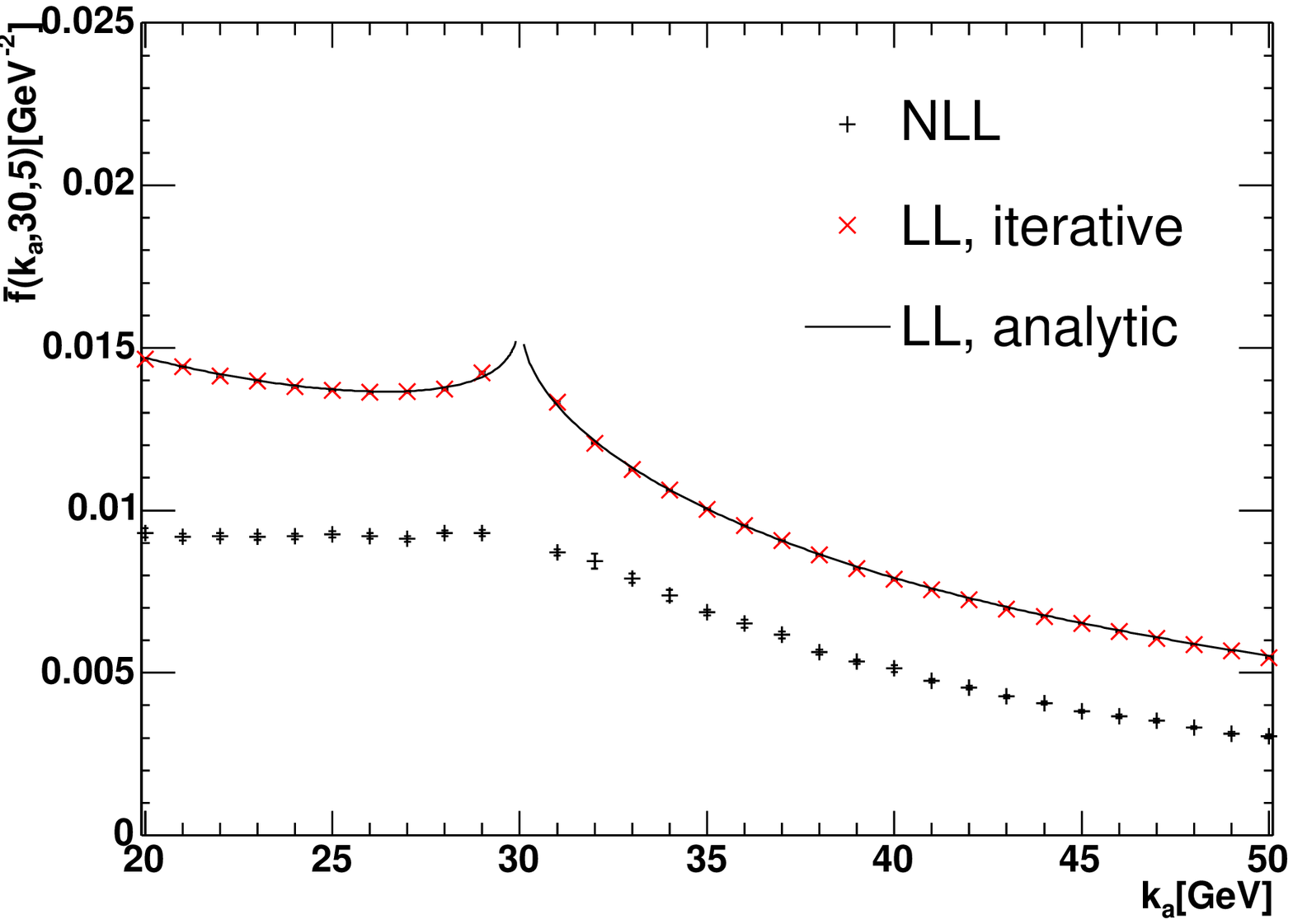}
  \caption{$k_a$ dependence of the LL and NLL gluon Green's function at $\mu=k_b=30$~GeV for two values of $\Delta$.}
  \label{fig:scale_scan}
\end{figure}
After describing the technicalities of our method of solution we are now
ready to show the behaviour of the NLL gluon Green's function. In
Fig.~\ref{fig:scale_scan} we have calculated $\bar f(k_a,k_b,\Delta)$ fixing
$k_b$ at 30 GeV and varying $k_a$ for the choice of parameter $\Delta=3$ and
$\Delta=5$. We start by noting the complete agreement between the iterative
and analytic results at LL. It is also interesting to notice how the angular
integrated NLL gluon Green's function evolves from being strongly peaked in
the region $k_a\simeq k_b$ for small values of $\Delta$ to being more flat in
$k_a$ for larger values.  This behaviour is to be expected, since the BFKL
equation can be reformulated as a differential equation in $\Delta$, with
$\delta^{(2)} \left({\bf k}_a - {\bf k}_b\right)$ as the boundary condition
at $\Delta=0$.  In agreement with this statement we obtain from
Eq.~(\ref{ours}) that $f \left({\bf k}_a,{\bf k}_b, \Delta = 0 \right) =
\delta^{(2)} \left({\bf k}_a - {\bf k}_b\right)$. By comparing the plots in
Fig.~\ref{fig:scale_scan} we also see how the gluon Green's function evolves
from the boundary condition as $\Delta$ increases.

\begin{figure}[tbp]
  \centering
  \epsfig{width=10cm,file=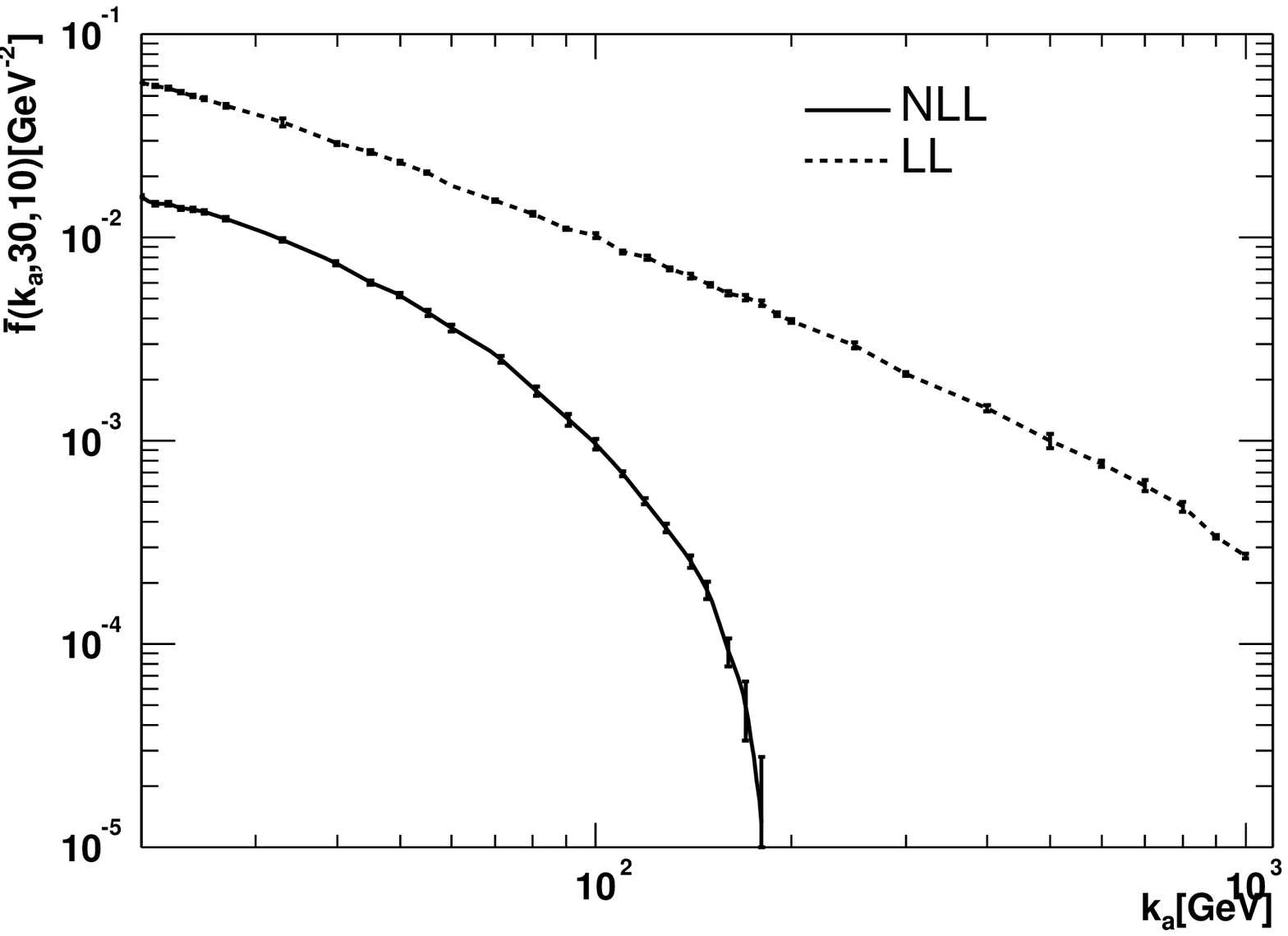}
  \caption{$k_a$ dependence of the LL and NLL gluon Green's function at $\mu=k_b=30$~GeV for $\Delta = 10$ and a large range in the $k_a/k_b$ ratio.}
  \label{fig:scale_scan_delta10}
\end{figure}

Although the normalisation of the gluon Green's function is different between
the LL and NLL cases, and indeed changes with $\Delta$, the shape with
respect to $k_a$ does not change too much in the $k_a \simeq k_b$ region.
This behaviour is different when $k_a \gg k_b$, as can be seen in
Fig.~\ref{fig:scale_scan_delta10}. In this region the gluon Green's function
is much suppressed at NLL compared to the LL case. Similar results were
obtained in Ref.~\cite{CCSS} when the solution to the BFKL equation was
studied using the Mellin transform of the angular averaged kernel, and a
similar treatment of the running coupling terms as in our function $\xi$ was
considered, i.e. the $\beta_0$ terms not resummed into $\alpha_s$. In
Ref.~\cite{CCSS} it was also shown that the resummation of these $\beta_0$
terms improves the behaviour of the gluon Green's function. We will come back
to this point in a forthcoming publication.

In principle, it would be interesting to establish whether the Green's
function, as obtained in the present approach, exhibits the exponentially
suppressed, oscillatory behaviour predicted in Ref.~\cite{Douglas}. For the
accessible range of rapidities in our present numerical study it was not
possible to verify this effect which would take place at large values of the
$k_a/k_b$ ratio and $\Delta$.

We now proceed to the study of the dependence on $\Delta$ in the next Section.

\subsection{Dependence of the Gluon Green's Function on $\Delta$}
\label{sec:depend-gluon-greens-1}
\begin{figure}[tbp]
  \centering
  \epsfig{width=10cm,file=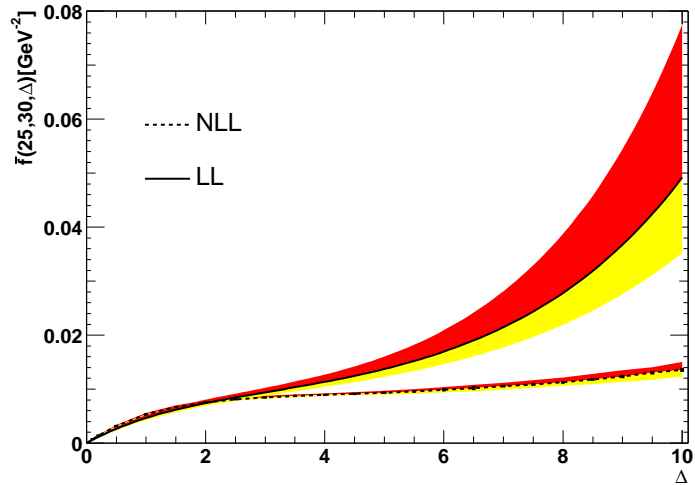}
  \caption{$\Delta$--dependence of the NLL gluon Green's function evaluated
    for $k_a=25$~GeV and $k_b=30$~GeV. The central lines for both the LL and
    NLL result are obtained by choosing the renormalisation scale $\mu=k_b$.
    The limits of the two upper (red) bands correspond to a choice of
    renormalisation scale $\mu=k_b/2$, while for the bound of the two lower
    (yellow) bands, the renormalisation scale is $\mu=2k_b$.}
  \label{fig:Delta_scan}
\end{figure}
In Fig.~\ref{fig:Delta_scan} we have plotted the evolution of $\bar
f(\mathbf{k}_a,\mathbf{k}_b,\Delta)$ with $\Delta$ for a specific choice of
$k_a$ and $k_b$, both for the LL and NLL case. We see that for this choice of
momenta, there is a significant difference in the $\Delta$--dependence from
about $\Delta=3$. With our choice of renormalisation scale of $\mu=k_b$, the
gluon Green's function rises slower with $\Delta$ at NLL than at LL.  For the
range in $\Delta$ studied in this paper no inestability in the growth of the
gluon Green's function has been found. It is known from previous
investigations in the literature~\cite{CCSS} that a correct resummation of
the $\beta_0$ terms into the running of the coupling eliminates this possible
problem. We will investigate this further in the future.

We study a renormalisation scale dependence by, in Fig.~\ref{fig:Delta_scan},
including the results for the choices $\mu=k_b/2$ and $\mu=2k_b$. At LL,
where the coupling does not run, we can still estimate a similar dependence
to that at NLL by simply choosing the fixed value of the coupling at LL to be
$\alpha_s(\mu=k_b)$, $\alpha_s(\mu = k_b /2)$ and $\alpha_s(\mu = 2 k_b)$.
This generates the band around the LL result in Fig.~\ref{fig:Delta_scan}.
The upper (red) bands are obtained for $\mu<k_b$, while the lower (yellow)
bands correspond to $\mu>k_b$.  This renormalisation scale dependence is very
big at LL and growing with $\Delta$.  For the particular selection of
parameters, the scale uncertainty is drastically reduced at NLL, with a slow
growth with $\Delta$.  This last feature has its origin in that as $\Delta$
increases, more powers of $\alpha_s$ are effectively resummed because there
is more phase space available for emission, as was seen in
Fig.~\ref{fig:ngplot}. The reduction in the uncertainty at NLL with respect
to the LL result shows how the predictive power of the theory has improved
with the inclusion of radiative corrections.  We will study this
renormalisation scale dependence in a toy cross--section below, which will
include an integration over a range of momenta. But before this, we first
analyse in the next Section the angular behaviour of the gluon Green's
function.

\subsection{Angular Dependence of the Gluon Green's Function}
\label{sec:angul-depend-gluon}
\begin{figure}[tbp]
  \centering
  \epsfig{width=10cm,file=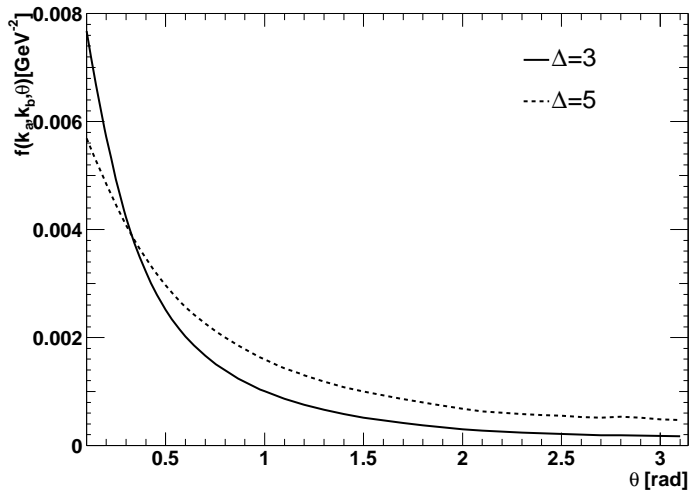}
  \caption{The dependence of the gluon Green's function at NLL on the angle
    between $\mathbf{k}_a$ ($k_a=25$~GeV) and $\mathbf{k}_b$ ($k_b=30$~GeV)
    for the choice of $\Delta=3$ and $\Delta=5$. The renormalisation point is
    chosen at $\mu=k_b$.}
  \label{fig:angle_scan}
\end{figure}

One of the advantages of our method of solution is that we can perform
studies of angular correlations in the NLL BFKL gluon Green's function. 
For example, in Fig.~\ref{fig:angle_scan} we show the dependence of
$f({k}_a,{k}_b,\theta,\Delta)$ defined in Eq.~(\ref{aver}) for $k_a = 25$~GeV
and $k_b=30$~GeV, and the two values $\Delta=3$ and $\Delta=5$. Again we see
that for smaller $\Delta$, the transverse momenta $\mathbf{k}_a$ and
$\mathbf{k}_b$ are strongly correlated, with the bulk of the contributions to
the gluon Green's function coming from the region of small angles between the
transverse momenta $\mathbf{k}_a, \mathbf{k}_b$. We have included an
investigation of the angular correlations of a toy cross section in
Section~\ref{sec:toy-cross-section}.

\subsection{Toy Cross Section}
\label{sec:toy-cross-section}
We now proceed to the study of the following quantity:
\begin{equation}
\label{eqS}
  S(\Delta)=\int_{k_a>30\mathrm{GeV}} 
\frac{d^2 {\bf k}_a}{{\bf k}_a^2}
\int_{k_b>30\mathrm{GeV}}  \frac{d^2 {\bf k}_b}{{\bf k}_b^2} 
f \left({\bf k}_a,{\bf k}_b, \Delta\right),
\end{equation}
which, when the gluon Green's function is evaluated at LL, is proportional to
the cross section at this accuracy. We therefore call it a ``toy'' cross
section. A more complete study would require the use of the full NLL impact
factors for different physical processes, a calculation which is out of the
scope of this paper. Although the accuracy of the impact factors, LL, does
not match that of the gluon Green's function, NLL, the behaviour of
$S(\Delta)$ in Eq.~(\ref{eqS}) with $\Delta$ will still give an indication of
the intercept we can expect at NLL.

\begin{figure}[tbp]
  \centering
  \epsfig{width=10cm,file=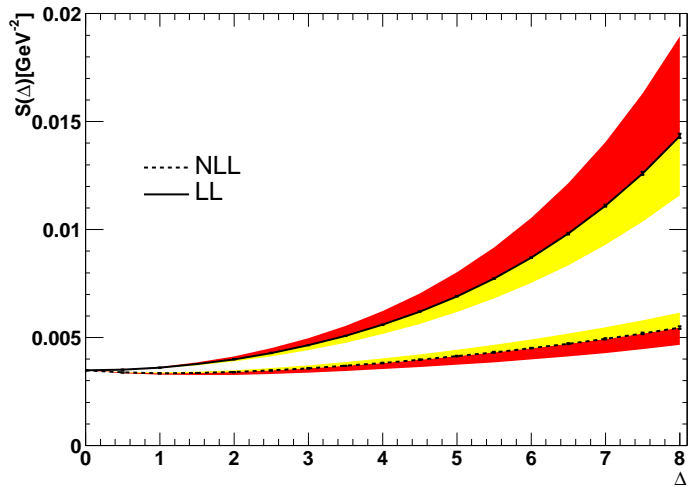}
  \caption{$S(\Delta)$ calculated at LL and NLL, including the
    renormalisation scale dependence as detailed in the text.}
  \label{fig:Integrated_Delta_scan}
\end{figure}
The evolution of the toy cross section with $\Delta$ is shown in
Fig.~\ref{fig:Integrated_Delta_scan} both for the LL and the NLL case. This
plot is consistent with the fact that the NLL corrections decrease the
intercept of cross sections.  With the intention to estimate the
renormalisation scale dependence in the LL curve, instead of keeping the
strong coupling fixed at, for example its value at the lower integration
limits $\alpha_s (\mu = 30 \,{\rm GeV}) \simeq 0.14$, we fix it at $\alpha_s
(\mu = k_b)$ as we do in the NLL calculation and in our study of the gluon
Green's function in
Sections~\ref{sec:depend-gluon-greens}--~\ref{sec:angul-depend-gluon}.  Again
we see, as in Fig.~\ref{fig:Delta_scan}, that the renormalisation scale
uncertainty is reduced when going from the LL to the NLL curve for our
particular choice of $\mu$. When the renormalisation scale is chosen to be
larger, the effective value of the coupling is reduced and the LL rise
decreases, while the size of the NLL corrections, suppressed by one power of
$\alpha_s$, diminishes. The overall effect is that the difference between the
LL and NLL curves is now smaller than for a lower choice of renormalisation
scale.  The initial decrease in $S(\Delta)$ in going beyond the LL
approximation was already predicted in models adding running coupling effects
to the LL evolution~\cite{LLenergy}.

\begin{figure}[tbp]
  \centering \epsfig{width=10cm,file=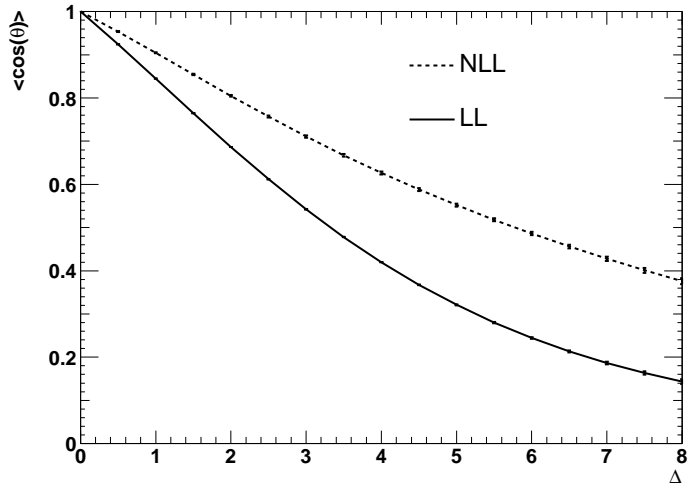}
  \caption{Average value of $\cos(\theta)$ for $S(\Delta)$ at LL and NLL as a
    function of $\Delta$. At $\Delta=0$ the two transverse vectors
    $\mathbf{k}_a$, $\mathbf{k}_b$ are completely correlated, whereas the
    BFKL evolution leads to a growing decorrelation with $\Delta$, both at LL
    and NLL. This decorrelation is less at NLL than at LL.}
  \label{fig:avgcostheta}
\end{figure}

Finally, and in order to illustrate the feasibility of our calculations, in
Fig.~\ref{fig:avgcostheta} we have plotted the average value of
$\cos(\theta)$, with $\theta$ being the angle between $\mathbf{k}_a$ and
$\mathbf{k}_b$, for the toy cross section as a function of $\Delta$. We see
that as $\Delta$ increases, $\mathbf{k}_a$ and $\mathbf{k}_b$ become
increasingly decorrelated, and we see that the effect is bigger at LL than at
NLL. 

\section{Conclusions}
\label{sec:Conclusions}
In this study we have presented a first analysis of the behaviour of the NLL
gluon Green's function as obtained from a numerical implementation of the
method proposed in Ref.~\cite{Andersen:2003an}. With the main purpose of
showing the feasibility of our method we have taken a particular choice of
renormalisation scale in the ${\overline {\rm MS}}$ renormalisation scheme.
Other choices and schemes can be considered, and work is in progress to study
them. We have shown that the convergence properties are well understood, and
in particular we have demonstrated the infrared finiteness of our solution.
We have also presented results on the evolution and angular dependence of the
gluon Green's function and a toy cross section. The intercept obtained from
this procedure decreases at NLL with respect to the one obtained at LL, a
trend which is in agreement with results in the literature~\cite{NLLpapers}.
The magnitude of the change when going from LL to NLL depends on the choice
of renormalisation scale and further studies are needed to draw stronger
conclusions.

\noindent {\bf Acknowledgements}
We would like to thank Victor Fadin, Stefan Gieseke, Gregory Korchemsky and
Lev Lipatov for very useful discussions, the CERN Theory Division for
hospitality, and the IPPP, University of Durham, for use of computer resources.
A.S.V.~thanks the II.~Institut f{\" u}r Theoretische Physik at the University
of Hamburg and the Laboratoire de Physique Th{\' e}orique at Universit{\'e}
Paris XI for hospitality, and acknowledges the support of PPARC (Postdoctoral
Fellowship: PPA/P/S/1999/00446).

\end{document}